\documentclass{emulateapj}
\usepackage{rotating}

\newcommand{\HST}{{\em HST}}
\newcommand{\combo}{{\sc combo-17}}

\newcommand{\aegis}{{\sc aegis}}
\newcommand{\slacs}{{\sc slacs}}
\newcommand{\cosmos}{{\sc cosmos}}
\newcommand{\gems}{{\sc gems}}
\newcommand{\goods}{{\sc goods}}
\newcommand{\galfit}{{\sc Galfit}}

\newdimen\hdsize
\shorttitle{Lens Candidates in the E-CDFS}
\shortauthors{More et al.}

\begin{document}

\title{Gravitational lens candidates in the E-CDFS}

\author{A.\ More\altaffilmark{1,2}, K.\ Jahnke\altaffilmark{3},
  S.\ More\altaffilmark{1}, 
  A.\ Gallazzi\altaffilmark{3,4},
  E.\ F.\ Bell\altaffilmark{5},
  M.\ Barden\altaffilmark{6}, 
  B.\ H\"au{\ss}ler\altaffilmark{7}, 
}

 \altaffiltext{1}{Kavli Institute for Cosmological Physics, University of
 Chicago, 5640 S. Ellis Ave., Chicago-60637, IL, USA;
 anupreeta@kicp.uchicago.edu}
 \altaffiltext{2}{Laboratoire d'Astrophysique de Marseille, 38 rue Frederic
 Joliot Curie, 13013 Marseille, France}
 \altaffiltext{3}{Max-Planck-Institut f\"ur Astronomie, K\"onigstuhl 17,
  D-69117 Heidelberg, Germany}
 \altaffiltext{4} {Dark Cosmology Centre, Niels Bohr Institute, Juliane Maries
 Vej 30, 2100 Copenhagen, Denmark}
 \altaffiltext{5}{Dept. of Astronomy, University of Michigan, 500 Church street,
 Ann Arbor, MI 48109, USA}
 \altaffiltext{6}{Universit\"at Innsbruck, Technikerstrasse 25, A-6020
  Innsbruck, Austria}
 \altaffiltext{7}{School of Physics and Astronomy, U. of Nottingham, NG7 2RD, UK }

\begin{abstract}

We report ten lens candidates in the E-CDFS from the \gems \ survey.
Nine of the systems are new detections and only one of the candidates is a known
lens system. For the most promising five systems including the known
lens system, we present results from preliminary lens mass modelling, which
tests if the candidates are plausible lens systems. Photometric redshifts of the
candidate lens galaxies are obtained from the \combo\ galaxy catalog. Stellar
masses of the candidate lens galaxies within the Einstein radius are obtained by
using the $z$-band luminosity and the $V-z$ color-based stellar mass-to-light
ratios. As expected, the lensing masses are found to be larger than the stellar
masses of the candidate lens galaxies. These candidates have similar dark matter
fractions as compared to lenses in \slacs\ and \cosmos.  They also roughly
follow the halo mass--stellar mass relation predicted by the subhalo abundance
matching technique. One of the candidate lens galaxies qualifies as a LIRG and
may not be a true lens because the arc-like feature in the system is likely to
be an active region of star formation in the candidate lens galaxy. Amongst the five best
candidates, one is a confirmed lens system, one is a likely lens system, two are
less likely to be lenses and the status of one of the candidates is ambiguous.
Spectroscopic follow-up of these systems is still required to confirm lensing
and/or for more accurate determination of the lens masses and mass density
profiles. 

\end{abstract}


%%%%%%%%%%%%%%%%%%%%%%%%%%%%%%%%%%%%%%%%%%%%%%%%%%%%%%%%%%%%%%%%%%%%%%%%%%%%%%
%
\section{INTRODUCTION}
 
Within the current standard cosmological model of the Universe, the
gravitationally dominant component of matter is dark. Hence, dark matter governs
the formation and evolution of the luminous matter like galaxies in the
Universe.  Gravitational lensing has proved to be a promising technique in the
past few decades allowing us not only to measure the mass of dark matter halos
which host luminous galaxies in their centres but also to probe the underlying
dark matter distribution of the lensing halo. 

Lensing has been used, for example, to constrain the slopes of density profiles
in the central regions of galaxies \citep[e.g.,][]{koopmans06,more08} and
clusters \citep[e.g.,][]{sand02,kneib03}, and to constrain cosmological
parameters like the Hubble constant to a better accuracy
\citep[e.g.,][]{fassnacht02,coles08} and the cosmological constant
\citep[e.g.,][]{oguri08}. Lensing surveys complemented with Hubble Space
Telescope (\HST) observations have explored various aspects of galaxy formation
and evolution like studying the interstellar medium of lens galaxies
\citep{falco99}, surface brightness evolution \citep{kochanek00}, mass density
profile evolution \citep{koopmans06}, and stellar populations \citep{treu06} of
early-type galaxies.

Among the large-area surveys at high spatial resolution, the \gems\ (Galaxy
Evolution from Morphologies and SEDs, \citealt{rix04}) survey in the Extended
Chandra Deep Field South (E-CDFS) is the largest contiguous \HST\ survey with
color imaging data to date. \gems\ was initiated to study the evolution of
normal and active galaxies out to high redshifts.  \gems\ observed 0.22~sq.~deg.
inside the E-CDFS in the optical F606W (=$V$) and F850LP (=$z$) filter band
passes with one orbit per pointing.  In the center, it contains the \goods\
south area. Of $\sim$60\,000 objects in the field, $\sim$14\,000 are galaxies
down to $R=24$ with photometric redshifts from the \combo\ survey.  The spectral
energy distributions (SEDs) available from \combo, besides allowing accurate
photometric redshift estimates, allow classification into various galaxy types,
including active nuclei \citep{wolf04}. Stellar mass estimates for all the
galaxies are also available from SED fitting \citep{borc06}. 

The depth and relatively large volume of the GEMS survey is apt for finding
incidences of strong gravitational lens systems.  For example, the strong lens
survey from All-wavelength Extended Groth strip International Survey (\aegis)
which has comparable depth and survey area to \gems, has found a sample of
three lenses \citep{moustakas07}. Also, the strong lens survey from COSMOlogical
evolution Survey (\cosmos) which has a much larger survey area in a single band
with \HST, find about 20 good lens candidates \citep{faure09}. An unbiased
sample of lenses can be used to constrain the faint end of the background source
luminosity function, to measure the optical depth for lensing, to study the dark
matter fraction in galaxies and its evolution, and to study the connection
between dark matter halos and galaxy properties.

In this paper, we present a sample of \gems\ lens candidates in the E-CDFS. The
structure of the paper is as follows. In section 2, we calculate the expected
number of lenses and explain the steps involved in the selection of the lens
candidates. In section 3, we describe the decomposition of \HST\ images into
candidate lensing galaxy and candidate lensed images. The decomposition is
carried out by using the two-dimensional surface brightness modelling program
\galfit\ \citep{peng10}.  We model the candidate lens systems as gravitational
lenses and check whether the derived model parameters are consistent with a
gravitational lens case. If the image configuration is not due to gravitational
lensing, it is likely that no sensible lens model can reproduce the lensed image
configuration.  We present results of the mass modelling in section 4.  In
section 5, we compute the stellar mass of the candidate lensing galaxy, using
color-based stellar mass-to-light ratio (MLR) and compare with the lensing
masses as a sanity check.  Furthermore, we compare the sample properties with
those in the literature, and discuss the probability of the candidates being
lens systems. We end with a brief summary in section 6. We use the following
cosmological parameters in our calculations: $\Omega_0=0.3$,
$\Omega_{\Lambda}=0.7$, and H$_0=70$~km~s$^{-1}$~Mpc$^{-1}$.
%%%%%%%%%%%%%%%%%%%%%%%%%%%%%%%%%%%%%%%%%%%%%%%%%%%%%%%%%%%%%%%%%%%%%%%%%%%%%%
%
\section{GEMS LENS SAMPLE}
\subsection{Expected Number of Lenses}
\label{enl}
The number of lenses expected to be found in a given survey depends upon the
survey area, the source number density, $n_{\rm s}$, and the optical depth for
lensing, $\tau$. The number of lenses per unit steradian, $n_{\rm l}$ is given
by
\begin{equation}
\label{eq:nldens}
n_{\rm l}=\int_{z_{\rm s1}}^{z_{\rm s2}} \tau(z_{\rm s},m_{\rm{lim}}) \,n_{\rm
s}(z_{\rm s},m_{\rm{lim}}) dz_{\rm s}\,,
\end{equation}
where $z_{\rm s}$ is the source redshift and $m_{\rm{lim}}$ is the 
intrinsic source limiting magnitude. We estimated the expected number of
lenses in the \gems \ survey by using optical depths from Table~2 of
\cite{faure09}. These optical depths were estimated by integrating the lensing
cross-section of halos in the Millennium Simulation. We calculated $n_{\rm s}$
by following Eq.~1$-$Eq.~3 of \cite{faure09} which requires the distribution of source redshifts
and source counts. We adopt the source
redshift distribution given by 
\begin{equation}
\label{zspdf}
p(z_{\rm s},m_{\rm lim}) = \frac{\beta\,z_{\rm s}^2}{\Gamma(3/\beta)\,z_0^3} \, {\rm exp}\left(-\left[\frac{z_{\rm s}}{z_0}\right]^\beta\right)\,,
\end{equation}
where $\beta=3/2$ and $z_0=0.13 \, m_{\rm lim}-2.2$ and use the cumulative source counts
distribution from the F814W($I$) band data of the Hubble Deep Field South
\citep[HDF-S,][]{casertano00}. The $V$ band limiting 
magnitude of \gems \ is used to calculate a corresponding $I$ band limiting magnitude.
The HDF-S sources follow the relation $V-0.49 \simeq I$ and \gems\ has a $V$
band limiting magnitude of 28.25 \citep{rix04}. Thus, a corresponding
equivalent limiting magnitude in $I$ band for \gems \ would be 27.76.
We note that due to lensing magnification, sources which are
intrinsically fainter than the limiting magnitude might also fall into our
lens sample. This would mean that our calculation of expected no. of lenses based on sources
only brighter than limiting magnitude might be biased. However, we can
de-magnify the lensed images, using the magnification factors from the lens mass
models, to calculate the intrinsic magnitudes of respective
sources. These source magnitudes happen to lie well-above the limiting magnitude
used in our calculation. Therefore, assuming our lens mass models are correct,
our calculation of expected number of lenses is less likely to be biased due to this effect.
By integrating Eq.~\ref{eq:nldens} from $z_{\rm {s1}}=0$ to $z_{\rm {s2}}=3$, we
expect about eight lens systems in a survey area of 0.22~sq.~degrees, down to
the limiting magnitude. 

\smallskip

\subsection{Selection of Lens Candidates} 
We adopted the following approach to find gravitational lens
candidates in the GEMS field. As a first step, we selected all galaxies above a
stellar mass of $3\times10^{10}$ $M_{\sun}$ and redshift $z\le1.1$ from the
\combo\ catalog \citep{borc06} of 25~000 galaxies. 
Massive lens galaxies are generally more efficient lenses and produce larger
image separations which are easier to detect. These galaxies are also, generally,
the more luminous galaxies at each redshift. As a result, such galaxies have a
high signal-to-noise ratio (SNR). The cut at $z=1.1$ ensures that
galaxies with reliable redshifts are included in the sample. Also, this limit
corresponds  to the redshift at which the redder $z$-band filter moves beyond
the 4000\AA-break, so we would start to see pure UV images beyond this redshift. In
addition to a decrease in the SNR, more clumpy star forming structures show up
more prominently in such cases, increasing the chance for contamination of good
lens candidates with normal star-forming galaxies. These
limits helped us to narrow down our search to a massive subsample of galaxies
which are more likely to exhibit strong lensing signatures. The first step
leaves 1225 \combo\ galaxies of which 1054 are covered with ACS
imaging data by the \gems\ survey.

In the second step, we visually inspected the 1054 galaxies, looking at both
$V-z$ color composite images as well as the two bands individually in different
gray-scale stretches. This allowed us to see both the surface brightness
structures as well as color-differences, if any, between the image components.
We selected possible lens candidates based on the presence of a) multiple image
components or arc-like structures mimicking typical lensed image configurations,
and/or b) color differences to the main body of the central galaxy. We found 29
candidates at the end of this inspection with some being more likely lens
systems than others. Three people ranked these candidates visually from 1 being
the least likely to 4 being very likely candidates. A total of 19 candidates had
average ranks below 2. Of the remaining 10 candidates, five had average ranks
between 2 and 3 and five had ranks $>=3$. One of the top five candidates is a
confirmed lens system \citep[J033238-275653,][]{blakeslee04} with \combo\ ID
15422. We analyze this system along with the rest of the candidates. The
top five candidates with ranks $>=3$ are listed in Table~\ref{tab:sample} and
examined in the rest of this paper whereas the five low probability candidates
are described in the appendix.

%%%%%%%%%%%%%%%%%%%%%%%%%%%%%%%%%%%%%%%%%%%%%%%%%%%%%%%%%%%%%%%%%%%%%%%%%%%%%%

\section{Surface Brightness Modelling}
We use the publicly available software \galfit\ to model the
two-dimensional light distribution in images. \galfit\ is a very flexible
code and allows fits to the light distribution with a large choice of
radial surface brightness profiles, number of distinct components, choice
of PSFs. \galfit\ in its newest version \citep[V3,][]{peng10} allows
a more flexible analysis through the possibility of applying ``bending'' modes to components and adding a
banana-shaped bend to an otherwise elliptical shape. This facilitates
adapting the models of the source in the presence of gravitational arcs.

Generally, we model all systems with one or two components for the lensing
galaxy and up to four for the lensed images. The choice of component
numbers is not physically motivated but was adapted to describe the whole
system well enough so that the candidate lensing galaxy and the putative
lensed images can be well separated.  Testing was done iteratively from
repeated modelling runs with different numbers of components and different
degrees of freedom to each component e.g.\ bending modes switched on or
off. The final choice is slightly subjective but the solution regarding
the exact parameters, at the same time, is quite degenerate.  Since we do
not use output model parameters from \galfit\ itself, but only from the
extracted images, this procedure is very robust. We give the individual
number of components used for each object in Table~\ref{tab:sample}.

\begin{figure*} \begin{center}
\includegraphics[width=15cm]{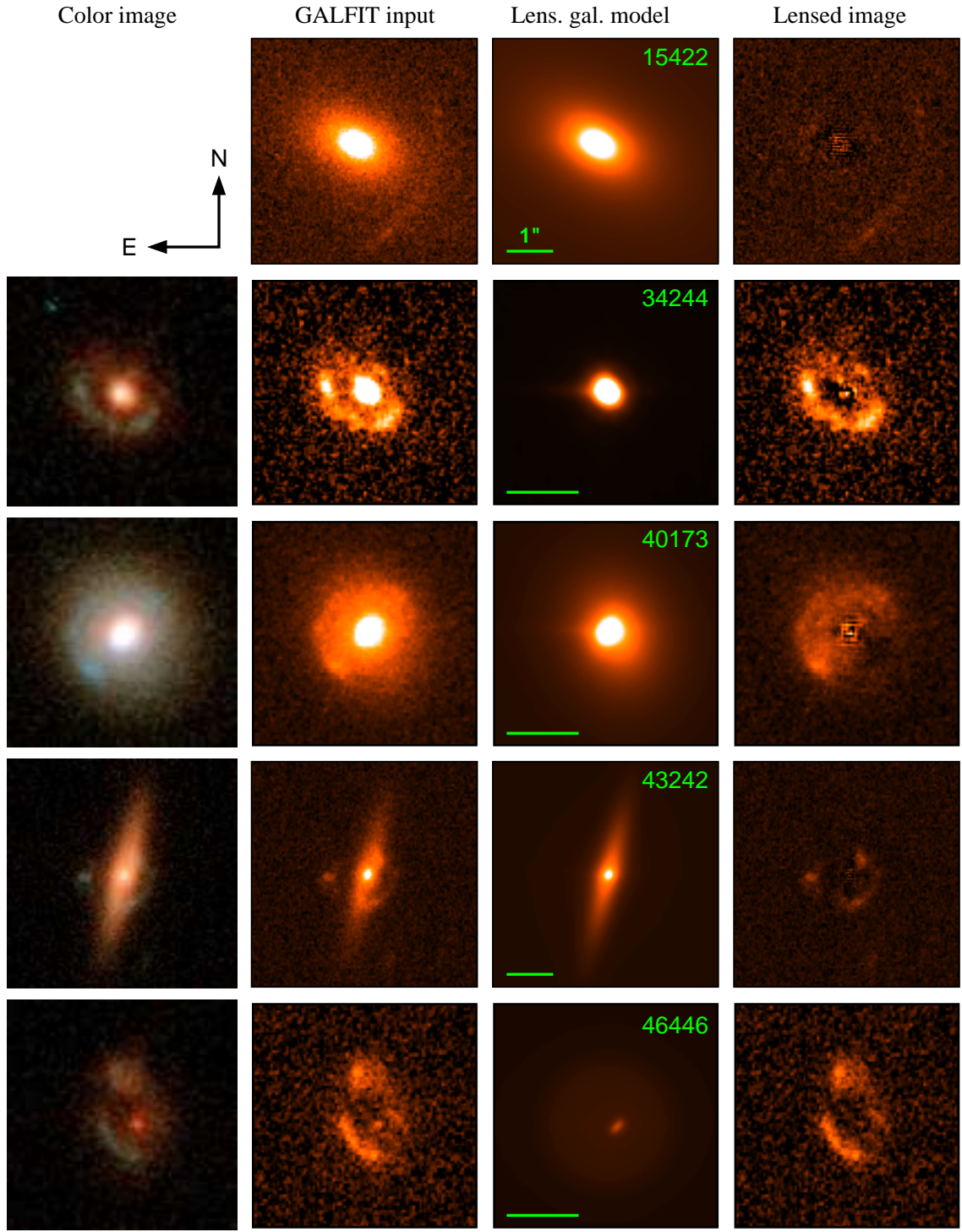} \end{center}
\caption{\label{fig:sample} 
Five lens candidates of the GEMS survey. The columns are (from left) $V$--$z$
color composite, $V$-band image ($z$-band for object 15422), surface brightness
model of the lens galaxy, and lens galaxy$-$subtracted residual image showing
emission from lensed images only. Markers show a scale of 1\arcsec.
The cutouts in the first and the fourth rows are 4.5$\times$4.5~arcsec$^2$
whereas the rest are 3.0$\times$3.0~arcsec$^2$.
} \end{figure*}

For four out of five objects, both $V$ and $z$ band images are available and we
model the deeper $V$ band images using the \galfit\ software. The object 15422
lies on GEMS tile \#44 for which the $V$ band is missing and thus, we revert to
modelling the $z$ band. Each row in Fig.~\ref{fig:sample} corresponds to one
candidate. From left, the columns show a composite $V$--$z$ color image of the
sample object, where available, as well as the \galfit\ modelling results: the
input $V$ (or $z$) band image, the surface brightness model for the lensing
galaxy, and the frame resulting from subtraction of the model showing only the
lensed images.

%%%%%%%%%%%%%%%%%%%%%%%%%%%%%%%%%%%%%%%%%%%%%%%%%%%%%%%%%%%%%%%%%%%%%%%%%%%%%%

\section{Lens Mass Modelling}
\subsection{Standard Mass Models}
The magnitude of deflection and distortion of a light ray bundle arriving at a
point in the lens plane from a background source depends upon the surface
density distribution of mass in the lens. To zeroth order, the surface mass
density distribution around galaxies follows a spherically symmetric singular
isothermal profile (SIS) given by
\begin{equation}
\label{eq:sigma1}
\Sigma (\xi) = \frac {\sigma^2}{2 \, \xi \, G} ~, 
\end{equation}
where $\sigma$ is the velocity dispersion of the dark matter particles, $\xi$ is
the physical separation in the lens plane and $G$ is the gravitational constant.
Dark matter halos are triaxial in nature. Therefore an elliptical mass
distribution is a more realistic description of the surface density
distribution. The expression that describes the surface density distribution for
a singular isothermal ellipsoid (SIE) with an axis ratio, q, is given by
\citet{kormann94},
\begin{equation}
\label{eq:sigma2}
\Sigma (\xi_1,\xi_2) = \frac {\sigma^2}{2 \, G} \sqrt{\frac{q}{q^2\xi_1^2+\xi_2^2}}\,. 
\end{equation}
Here, $\xi_1$ and $\xi_2$ denote the projected physical separations along
the major and minor axes of the distribution, respectively. The lensing
observables are sensitive to a quantity called convergence ($\kappa$)
which is defined as the ratio of the surface mass density distribution to
a critical surface density.  The critical surface density depends upon the
geometrical distances between the source, the lens, and the observer and is
given by $\Sigma_{\rm{crit}}~=~c^2 D_{\rm s} / (4 \pi G D_{\rm d} D_{\rm
  ds})$, where $c$ is the speed of light, $D_{\rm d}$ is the angular
diameter distance between the observer and the lens, $D_{\rm ds}$ is the
distance between the lens and the background source, and $D_{\rm s}$ is the
distance between the observer and the background source.

Using constraints from extended images is particularly helpful when 
the background source is to be reconstructed for further analysis. 

We use the software package {\sc gravlens} \citep{keeton01} to make lens mass
models for our candidates. The software package uses the image positions (and
optionally, their fluxes) and the lens position along with their error bars as
constraints and outputs a model that best fits the constraints. For the purpose
of estimating masses of the lens galaxies, use of the surface brightness peak
positions is usually sufficient. A significant improvement is not expected in
the estimate of the Einstein radius by using the extended surface brightness
information. This is postponed until deeper imaging is obtained especially 
for studying the background sources.

The convergence at a point ($x=\xi_1/D_{\rm d}$, $y=\xi_2/D_{\rm d}$) in the sky is defined
as\footnote{For further details, see eqs.~$3.24-3.26$ in the {\sc GRAVLENS}
manual.}
\begin{equation} \label{eq:kappa} \kappa(x,y) =\frac{\Sigma}{\Sigma_{crit}} =
\sqrt{\frac{1+q^2}{2}}\frac {b'} {2q \sqrt{x^2 + y^2/q^2}} \, , \end{equation}
where $b'$ is the critical radius (also known as the Einstein radius). As is clear
from Eqs. \ref{eq:sigma2} and \ref{eq:kappa}, the critical radius and the
velocity dispersion ($\sigma$) are related by

\begin{equation}
\label{vdisp}
 \sigma^2 = \sqrt{\frac{1+q^2}{2q}} \frac{c^2 }{4 \pi} \frac{D_s}{D_{ds}} \, b' \, .
\end{equation}

Integrating the surface mass density within the area enclosed by the tangential
critical curve\footnote{The tangential critical curve is an iso-density contour
within which the $\langle \kappa \rangle = 1$. It is the locus of points where
lensed images are infinitely magnified and are stretched in a tangential
direction.} yields the lens mass within the critical curve, $M_{\rm E}$, which
is given by

\begin{equation}
\label{mlens}
 M_{\rm E} = \frac{1+q^2}{2q}\frac{c^2}{4 G} \frac{D_d D_s}{D_{ds}} b'^2 \, .
\end{equation}

The first factor on the right hand side of Eq. \ref{mlens} becomes unity in the
case of a SIS profile (i.e., when $q=1$). In addition to the surface mass density
distribution, an external shear is often added to the lens model to account for
the tidal effects of the environment on the lensed image observables. The
position angles in the lens models are measured East of North such that North
is 0~deg and East is 90~deg. 

An SIS model has a total of 5 free parameters: the position $(x,y)$ of the lens,
its critical radius, and two parameters that describe the position of the source.
An SIE model has two additional parameters, the ellipticity (given by $1-q$) and
its position angle. The external shear introduces 2 more parameters, the shear
strength, and its position angle, to each of these models. A doubly imaged system
provides a total of 6 constraints: 4 constraints from the 2 image positions and
2 constraints from the lens position. In the absence of any constraints from the
flux ratios, we only use the SIS model so as to avoid an over-fit to the data.
Similarly, a quadruply imaged system will have a total of 10 constraints: 8
constraints from the 4 image positions and remaining 2 constraints come from the
lens position. This allows us to fit the data with SIS+shear and SIE+shear
models. In case of SIE+shear model, the external shear is often found to be
degenerate with the ellipticity.

%%%%%%%%%%%%%%%%%%%%%%%%%%%%%%%%%%%%%%%%%%%%%%%%
%
\subsection{Mass Modelling Results} 
\label{sect:massmod}
Using the above models, we present results from preliminary mass modelling of
the lens candidates. The aim is to test if simple mass models can reproduce the
configuration of lensed images with reasonable values of the model parameters.

Two kinds of image configurations are commonly seen in four-image lens systems,
namely, a cusp and a fold. In the former, three lensed images are highly
magnified and tend to merge whereas, in the latter, any of the adjacent two
images are highly magnified and tend to merge. We assume either of the above
image configurations (except for candidate 40173 which is probably doubly
imaged) for our mass models and describe the tests carried out on all of our
candidates. 

Throughout the modelling, we use peak positions of the images and the lens
galaxy as constraints for the mass models. The positions of the images were
visually determined at the peak of surface brightness (whenever possible)
whereas the peak position for the lens galaxy was chosen by fitting a 
Sersic profile. The errors on the positions of the images and the lens galaxy were
assumed to be $0.06$~arcsec and $0.03$~arcsec, respectively. Note that the pixel
scale is $0.03$~arcsec.

\begin{figure*}

\begin{center}
 \includegraphics[width=15cm]{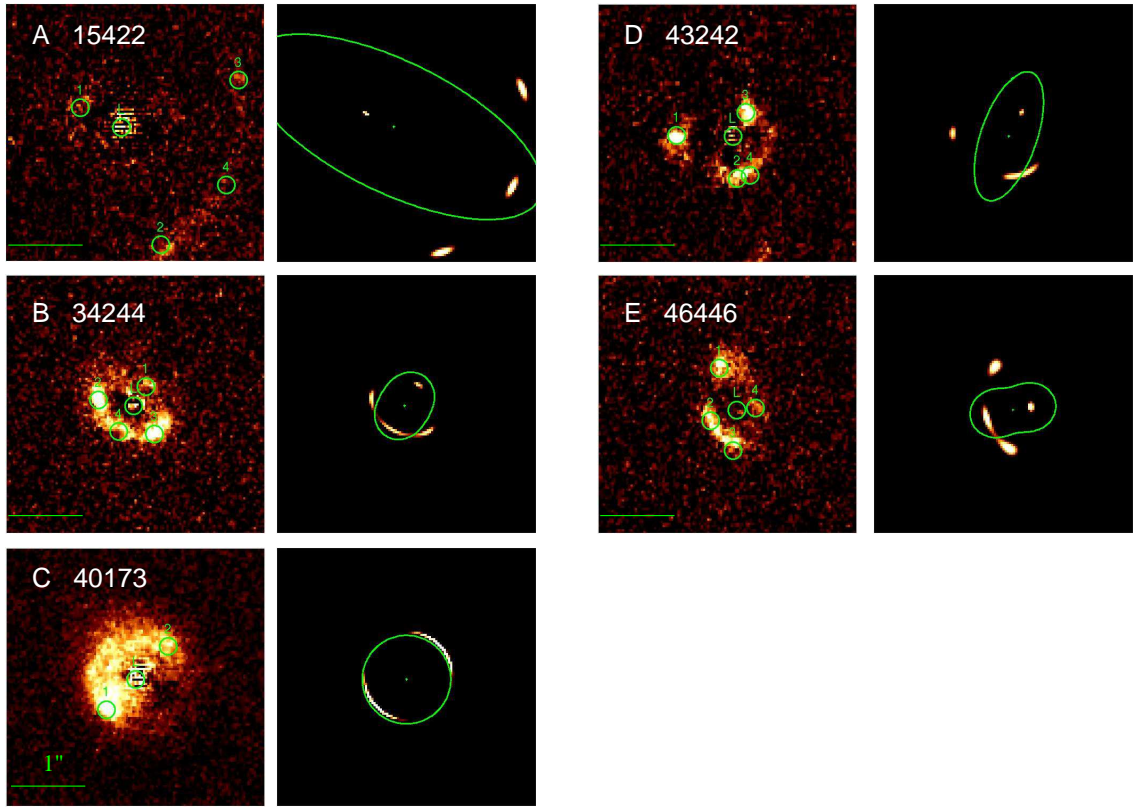}
\caption{\label{fig:modres}
Each candidate is represented by a pair of panels. The left panel in the pair
shows the model-predicted positions of lensed images and the lens galaxy
overlaid on the residual images. The right panel in the pair shows reconstructed
lensed images assuming an extended source and using the best-fitting lens model.
Also, critical curves are shown with green solid curves.}
\end{center}
\end{figure*}

In the following, we describe the mass modelling analysis for five of the best
candidates. The results of the mass models for each candidate are shown with a
pair of panels in Fig.~\ref{fig:modres}. Every left panel has the
model-predicted positions overlaid on the true lensed images with the foreground
galaxy subtracted. Every right panel shows a reconstructed lensed image
configuration assuming a Sersic model for the background source with
arbitrarily chosen values for the parameters and using the best-fitting model
for the lens galaxy. This is for illustrative purposes only. The solid green curves in the figure are the critical
curves corresponding to the best-fitting lens model.

{\bf 15422}: This is the only system from our sample which is known to be a
confirmed lens system \citep[hereafter, B04]{blakeslee04}. The lens galaxy has a spectroscopic
redshift of 0.62 and a photometric redhift of 2.4 is estimated for the
background source. The mass modelling of B04 suggests that the
velocity dispersion of the lens galaxy is $\sim$ 305 km s$^{-1}$.

For the sake of consistency and completeness, we make mass models with our data
and compare our modelling results with B04. In this system, one
can clearly see a thin elongated arc. A faint counter-image is usually expected
and can actually be seen in Fig.~\ref{fig:sample} where the lens galaxy has been
subtracted. This counter-arc is clearly visible in B04. We use
the peak positions of all four images as constraints to the mass models. 

We test an SIE mass model alone motivated by the elliptical light distribution
of the lens galaxy. A SIE model fits reasonably to this lens system with a reduced
$\chi^2$ of 3. The Einstein radius is found to be 1.20~arcsec. The mass model
suggests an ellipticity of 0.60 with a position angle of 64~deg. This is
consistent with the ellipticity and the position angle measured from the
distribution of light from the lens galaxy. The image positions predicted by
the SIE model are shown with circles in the left panel of A in
Fig.~\ref{fig:modres} overlaid on the lens-subtracted optical image. The right
panel of A in the figure shows a reconstructed lensed image assuming a source
lensed by the best-fitting SIE model. The green ellipse shows critical curves
of this model.

In conclusion, the SIE model constrained by the arc and counter-image provides a
reasonably good fit. A comparison of our modelling results with B04 is done in
Sect.~\ref{lme}.

\begin{table*}
%%\begin{sidewaystable}
\begin{center}
\caption{\gems\ strong lens candidates \label{tab:sample}}
\begin{tabular}{cccccccccccccc}
\tableline
\tableline
ID\tablenotemark{a}&
Tile\tablenotemark{b}&
RA\tablenotemark{c}&
DEC\tablenotemark{c}&
$V_\mathrm{tot}$\tablenotemark{d} &
$z_\mathrm{tot}$\tablenotemark{d} &
$z_\mathrm{l}$\tablenotemark{e}&
$z_\mathrm{lsp}$\tablenotemark{f}&
comp.\tablenotemark{g}&
log($M_{\rm{t}\ast})$\tablenotemark{h}&
log($M_{\ast}$)\tablenotemark{i}&
$b'$\tablenotemark{j}&
$q$\tablenotemark{k}&
log($M_{\rm{E}}$)\tablenotemark{l}\\
\tableline
 15422& 44& 03:32:38.21& --27:56:53.2&  --   & 19.20 & 0.58 & 0.62 &1+3& 11.4& 11.22$^{+0.22}_{-0.20}$ & 1.20  & 0.40 & 11.80$^{+0.15}_{-0.06}$\\
 34244& 94& 03:32:06.45& --27:47:28.6& 23.43 & 21.11 & 1.00 & 1.02 &1+4& 11.0& 10.82$^{+0.20}_{-0.29}$ & 0.40  & 0.48 & 11.04$^{+0.27}_{-0.11}$\\
 40173& 35& 03:33:19.45& --27:44:50.0& 20.86 & 19.67 & 0.44 & 0.42 &2+3& 10.7& 10.58$^{+0.13}_{-0.14}$ & 0.59  & 1.00 & 10.85$^{+0.09}_{-0.04}$\\
 43242& 45& 03:31:55.35& --27:43:23.5& 21.85 & 20.00 & 0.66 & 0.66 &2+3& 11.1& 11.04$^{+0.18}_{-0.14}$ & 0.57  & 0.60 & 11.08$^{+0.16}_{-0.07}$\\
 46446& 47& 03:31:35.94& --27:41:48.2& 23.60 & 21.70 & 0.88 &  --  &2+3& 10.6& 10.37$^{+0.23}_{-0.21}$ & 0.42* & 1.00 & 10.90$^{+0.23}_{-0.09}$\\
\tableline
\end{tabular}
\medskip
\begin{minipage}{\hdsize}

{\bf a}--ID from the \combo\ catalog. {\bf  b}--\gems\ tile number (1--63 \gems,
80--95 \goods\ region). {\bf c}--positions of candidates are in J2000 with units
of hrs:min:sec and deg:min:sec. {\bf d}--Apparent $V$=$F606W$ and $z$=$F814LP$
HST/ACS magnitudes of the whole system \citep{cald08}. {\bf e}--Photometric
redshift of lensing galaxy from \combo. {\bf f}--Spectroscopic redshift of lensing
galaxy (see Sect.~\ref{lme} for details). {\bf g}--Number of components used in \galfit\
composition for lensing galaxy + lensed images. {\bf h}--Total stellar mass of
the lens galaxy in units of $M_{\odot}$ from \cite{borc06}.  {\bf i}-- Stellar
mass of the lens galaxy inside the critical radius in the units of $M_{\odot}$.
{\bf j}--Einstein radius in arcsec corresponding to SIE model except * is
corresponding to SIS+shear model. {\bf k}-- axis ratio from the lens mass model
{\bf l}--Mass within the critical curve in units of $M_{\odot}$. The upper and
lower limits correspond to the 16$^{\rm{th}}$ and 84$^{\rm{th}}$ percentile of
the lens mass PDF.

\end{minipage}
\end{center}
%%\end{sidewaystable}
\end{table*}

{\bf 34244}: In this system, a cusp configuration is likely and a counter-image
is clearly detected. As before, we test three possible mass models namely, SIE,
SIS+shear and SIE+shear.

The SIE model suggests an ellipticity of 0.52 with a position angle of
$-31$~deg. The Einstein radius is found to be 0.40~arcsec. The model predicted
image positions overlaid on the lens-subtracted residual image are shown in left
panel of B in Fig.~\ref{fig:modres}. The right panel of B shows an artificial
source lensed with the best SIE model along with the critical curves. A
SIS+shear model indicates a high shear of about 15 per cent also at the position
angle of $-31$~deg and Einstein radius of 0.42~arcsec.

Both the ellipticity and shear could not be simultaneously constrained with the
image positions unless a prior on the ellipticity was used from the intensity
distribution. Thus, to test a SIE+shear model, an ellipticity of 0.36 with a
position angle of $-42$~deg was used from the light profile of the lens galaxy.
The best-fit model has a moderately high shear of 9 per cent with a position
angle of $-11$~deg.

All of the above models have reduced $\chi^2 \lesssim 1.5$ and hence, are
equally favorable. Using the constraints from the  image positions, the SIE, and
SIS+shear models suggest a position angle of $-31$~deg for the ellipticity in
the lens potential. On the other hand, the position angle of the intensity
distribution of the lens galaxy shows a significant offset from the model
predicted value indicating inconsistencies in the alignment of matter and light
distribution. It is unclear if any one of the above models is better than the
others due to lack of physically motivated values of the parameters. 

Here, the SIE model may be referred to as the current best-fitting
model. A subsequent investigation of the lens and its environment  might help in
testing whether the misalignment between the matter and light distribution is
physical. 

{\bf 40173}: Owing to the circular symmetry of the surface brightness
distribution, this candidate could be an example of a partial Einstein ring or a
double lens system. Since the lens-galaxy light subtraction may not be accurate,
the residual arm-like structure could be misleading. Instead, we choose a
two-image configuration. An SIS model, centered at the position of the lens
galaxy, can be tested using the peak positions of two images as constraints. An
Einstein radius of $0.59$~arcsec with a symmetric isothermal density profile
reproduces the image separation. Model predicted image positions and an ideal
source lensed by the mass model are shown in left and right panels of C in Fig.
\ref{fig:modres}, respectively.

However, fitting the image separation alone can not, necessarily, indicate
gravitational lensing in the system. Spectra of the lensed images will be
crucial in concluding whether blue arc-like feature is a site of star-formation
at the lens redshift or a lensed arc. Note that, if this indeed is a lens
system then size of the Einstein radius can give a very accurate measure of the
mass within the Einstein radius, irrespective of the assumed density profile.

{\bf 43242}: This lens system appears to be in a fold configuration. We assumed
that the two southernmost images are merging to form an arc-like feature.
With the positions of four images, following models are tested.  

Firstly, SIE and SIS+shear models are tested. Both the models fit equally well
with a reduced $\chi^2 \sim 2$. The Einstein radius of the lens galaxy is about
$ 0.57$~arcsec with an ellipticity of $0.4$ at a position angle of $-20$~deg or
alternatively, a shear with  $13$~per~cent strength at the same position angle.
The panels of D in Fig.~\ref{fig:modres} show the critical curves for an SIE
mass model on the right whereas the observed and model-predicted image positions
(green circles) on the left. The ellipticity of the bulge of the lens galaxy is
$0.32$ with a position angle of $-25$~deg which is consistent with the
ellipticity of the SIE model. Using this ellipticity as a constraint in a
SIE+shear model results in equally good fit. The shear for this model is $6$ per
cent with a position angle of $1$~deg. This implies that the bulge component
along with reasonable shear strength can reproduce the configuration of the
lensed images. 

Deep observations are certainly desirable to well-resolve the arc and high
surface brightness images which are contaminated due to the emission from
the lens galaxy. Furthermore, if this system hosts a lens then
spectroscopic redshifts will provide better constraints in measuring the
total mass of the late type lensing galaxy bulge. This would
be interesting since there are only few late type lenses known in the
literature \citep[e.g.,][]{feron09,sygnet10,dutton11}.

{\bf 46446}: This candidate indicates a fold configuration,
more obvious in the bottom-right residual image of Fig.~\ref{fig:sample}. We
model the lens system with standard isothermal models.

The SIS+shear and SIE models for this system suggest Einstein radii of
$0.42$~arcsec and $0.31$~arcsec for the lens galaxy, respectively. The reduced
$\chi^2$ for these models is $\sim 2$. The models need either high shear of
30~per~cent at a position angle of $-76$~deg or a high ellipticity of $0.85$ at
a position angle of $-79$~deg. The brightness profile fitting to the lens galaxy
also implies an unusually high ellipticity of $\sim 0.95$. However, the
extremely high shear is unexpected for a galaxy-scale lens system. Using priors
on the ellipticity from the light profile, a SIE+shear model is tested. The
parameter space of ellipticity and shear has a broad degeneracy resulting in a
range of equally favorable models. 

The left panel of E in Fig.~\ref{fig:modres} shows the predicted image positions
with green circles using SIS+shear model overlaid on the lens-subtracted image.
The right panel of E in Fig.~\ref{fig:modres} shows the critical curves and the
reconstructed image configuration assuming the best-fitting SIS+shear model
parameters which well-reproduce the positions of the observed lensed images. 

The high ellipticity of the lens from the mass model is inconsistent with the
morphology and optical properties of the central galaxy. The concentration of
light in the central regions of the lens galaxy does not suggest a high
concentration of mass which is required for it to be a lens.  In addition,
statistically, we expect more red galaxies with star-forming shells than red
lens galaxies. Therefore, it would be decisive to know the redshifts of the blue
features. 

%%%%%%%%%%%%%%%%%%%%%%%%%%%%%%%%%%%%%%%%%%%%%%%%%%%%%%%%%%%%%%%%%%%%%%%%%%%%%%

\section{DISCUSSION}
\subsection{Color-based Stellar Masses}
We compute the stellar mass $M_*$ inside of the critical curves (see Sect.
\ref{sect:massmod}) by weighting the luminosity in the $z$-band with the stellar
MLR as derived from the ($V-z$) color. According to \citet{gall09}, who used a
large Monte-Carlo library of star formation histories to assess the systematic
uncertainties in calculating stellar mass-to-light ratios, a blue rest-frame
optical color is a good proxy for the MLR. Dust attenuation,
metallicity, and age have degenerate effects on the MLR derived from colors and
on the luminosity. Incidentally, the effects act in opposite directions on MLR
and luminosity which cancel out to a good extent. As a result, the estimated
stellar mass is fairly robust against variations in these parameters, of which
it is not possible to have independent estimates.

We estimated the stellar masses for our lens galaxies with a Bayesian
statistical approach as outlined in \citet{gallazzi05}. We created a library of
model SEDs by convolving \citet{bruzual03} simple stellar population (SSP)
models of different metallicities, ranging from 20~per~cent to 2 times the solar
metallicity, with a Monte Carlo library of complex star formation histories
(SFH), and dust attenuations. We followed the prior distribution adopted by
\citet{salim05} and assumed an initial-mass-function (IMF) of
\citet{chabrier03}. The stellar mass of each model is computed by multiplying
the $ M_\ast/L_{\rm{F850LP}}$ of the redshifted model with the observed
luminosity $L_{\rm{F850LP}}$ of the lens galaxy.  Inputs for each target are the
observed $V$ and $z$ fluxes of the candidate lensing galaxy inside the critical
curve after \galfit-based subtraction of the lensed images as well as the
redshift of the lens galaxy. For each galaxy, we built the probability density
function (PDF) of $\log( M_\ast/M_\odot)$ by comparing the observed $V-z$ color
with the color of the models computed at the redshift of the galaxy. Only models
with formation age younger than the age of the Universe, at the redshift of the
galaxy, contribute to the PDF. The stellar mass of the galaxy is then estimated
as the median of the PDF with a 68\% confidence interval given by the
16$^{\rm{th}}-$84$^{\rm{th}}$ inter-percentile range. The results are listed in
Table~\ref{tab:sample}.

For the object 15422, we only have the flux at F850LP. Hence, no color
information can be obtained. Motivated by the S0-like morphology of this
galaxy, we assume that it has a rather smooth SFH and intermediate/old stellar
population. This translates into selecting models without a burst in the last
2~Gyr and with $4000$\AA$-$break values typical of S0 galaxies (roughly between
1.4 and 2, \citealt{poggianti08}), in addition to the constraint that the
formation age be younger than the age of the Universe at the galaxy redshift.
To estimate the stellar mass of this galaxy, instead of computing the PDF, we
simply take the median and percentiles of the distribution in $\log(
M_\ast/M_\odot)$ for the models satisfying these criteria.

%%%%%%%%%%%%%%%%%%%%%%%%%%%%%%%%%%%%%%%%%%%%%%%%
%
\subsection{Lensing Mass Estimates} \label{lme} 

The Einstein radii and the axis ratios of the best-fitting mass
models are listed in Table~\ref{tab:sample} along with the photometric and
spectroscopic (whenever available) redshifts of the lens galaxies. Spectroscopic
redshifts of the main galaxies were available for 15422 \citep{blakeslee04} and
34244 \citep{vanzella05}. Furthermore, as a part of Arizona CDFS Environment
Survey (ACES) spectroscopic redshifts have been measured for the main galaxies
in 40173 and 43242 (Cooper et al. 2011, in prep). Thus, we use the photometric
redshift for 46446 only although we note that any flux contamination from the
putative lensed features might bias the photometric redshift estimate and hence,
the lens mass. Any quantitative estimation of this bias would require tests with
simulations which is beyond the scope of this paper. Since the redshifts of the lensed
sources are not known, we use the PDF of the redshifts of the background
sources\footnote{We note that B04 provide a photometric redshift for the
background source in 15422. However, we use the source redshift PDF as a sanity
check of our analysis and compare our lens mass estimate to that of B04.} (see
Eq.~\ref{zspdf} in Sect.~\ref{enl}. Thus, masses of the lens galaxies within the
Einstein radii are calculated by substituting above described parameters in
Eq.~\ref{mlens}. The median masses are given as $M_{\rm E}$ in
Table~\ref{tab:sample}. The error bars reflect the 16$^{\rm{th}}$ and
84$^{\rm{th}}$ percentile of the PDF of mass, given the lens model parameters
and the source redshift PDF.

Using the Einstein radius, lens and source redshifts of B04 in case of 15422, we
find that their lens mass within the Einstein radius is log$(M_{\rm
E})=11.86~M_{\odot}$. This mass is consistent with our estimate of log$(M_{\rm
E})=11.80^{+0.15}_{-0.06}~M_{\odot}$ from our mass model and assuming a PDF for
source redshift instead (see Table~\ref{tab:sample}). In the case of 40173, we note
that there is a hint of an inner arc-like feature from another source which may
or may not be at the same redshift as the source lensed into the outer arc.
Given the quality of the image it is difficult to say which of the blue features
in the inner and outer arcs correspond to which sources. Nevertheless, if we
assume that the outer arc does have a counter-arc on the inner side, then the
fractional change in the masses is about 0.1 dex which is within the uncertainty
arising from the lack of knowledge of the source redshift. 
%
%%%%%%%%%%%%%%%%%%%%%%%%%%%%%%%%%%%%%%%%%%%%%%%%
%
\subsection{Comparison of Lensing and Stellar Masses} \label{sect:masscomp}
Table~\ref{tab:sample} gives the mass estimates obtained from lens mass
modelling and stellar masses within the critical curve of each system.
Comparison of the two masses shows that lens mass estimates are consistently
higher than the respective stellar masses of all the candidates. This conforms
to the scenario that lensing measures total projected mass (dark matter and
luminous matter) and should be larger than the stellar mass, which is sensitive
to the luminous mass alone, within the same aperture. 

\begin{figure} \begin{center}
\includegraphics[width=8.5cm]{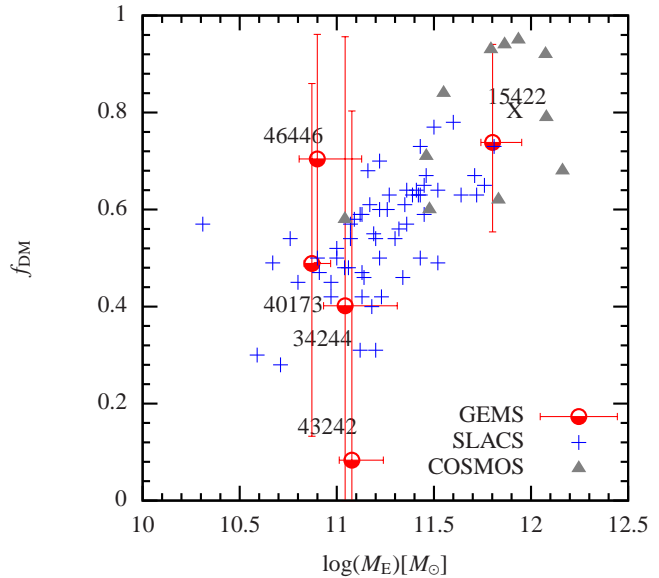} \end{center}
\caption{\label{fig:fdm} 
Projected DM fraction as a function of lens mass ($M_{\rm E}$). The
projected DM fractions in \gems\ candidates are similar to those found in the
lenses from the \slacs\ and \cosmos\ surveys. The label ``X" shows the location of 15422
using the lens mass from B04.}
\end{figure}

Early-type galaxies happen to lie on a Fundamental Plane (FP) defined by the
surface brightness, the effective radius, and the velocity dispersion.
However, observationally the early-type galaxies are found to
deviate from the naive L $\propto \sigma^2 R$ expectation and this deviation is referred to as the
tilt of the FP. The origin of the tilt is thought to be due primarily to
the content of dark matter in the early-type galaxies \citep[e.g.,][although
variations in stellar MLR and structural non-homology can affect the tilt as
well, possibly to a minor degree]{padma04}. Projected dark matter (DM) fractions
for lens galaxy samples have been calculated in surveys like \slacs\ \citep{auger10}
and \cosmos\ \citep{faure10}. The projected DM fraction is found to be
positively correlated with the lens mass in early-type lensing galaxies as shown
in Fig.~\ref{fig:fdm}. The projected DM fraction ($f_{\rm DM}=1 -
M_{\ast}/M_{\rm E}$) is calculated for all the \gems\ candidates and shown as a
function of the lens mass ($M_{\rm E}$) in Fig.~\ref{fig:fdm}.  For
15422, we also estimate the $f_{\rm DM}$ using the lens mass of B04 and stellar
mass from Table~\ref{tab:sample}. This is denoted by label ``X" in
Fig.~\ref{fig:fdm} and is consistent with our estimate within the
uncertainties. Comparison with the \slacs\ and \cosmos\ lenses shows that the
\gems \ candidates roughly follow the same correlation. A more conclusive result
can be presented when the background source redshifts are measured which will
accurately determine the lens masses.

We further checked where do the candidate lens galaxies lie on the halo mass$-$stellar mass
plane. To test the above, we need to compare the mass of the halo
with the total stellar masses. We calculated the velocity dispersion ($\sigma$)
using the Einstein radius and Eq.~\ref{vdisp}.  We used the $\sigma$ to
calculate the virial mass ($M_{\rm{vir}}$) for each of the lens galaxies
following the relation given by \citet{bryan98},
\begin{equation}
M_{{\rm vir}}= \frac{4~\sigma^3}{H(z) G \sqrt{\Delta_{\rm c}}} .
\end{equation}
Here, $H(z)$ is the redshift dependent Hubble parameter and $G$ is the
gravitational constant. The virial mass is defined to be the mass of
the halo within a radius that encloses a density contrast of $\Delta_{\rm c}$
with respect to the critical density of the Universe. We used the
fitting function provided by \citet{bryan98} to calculate $\Delta_{\rm c}$ as a
function of the cosmological parameters and the redshift. 

\begin{figure}
\begin{center}
\includegraphics[scale=0.90]{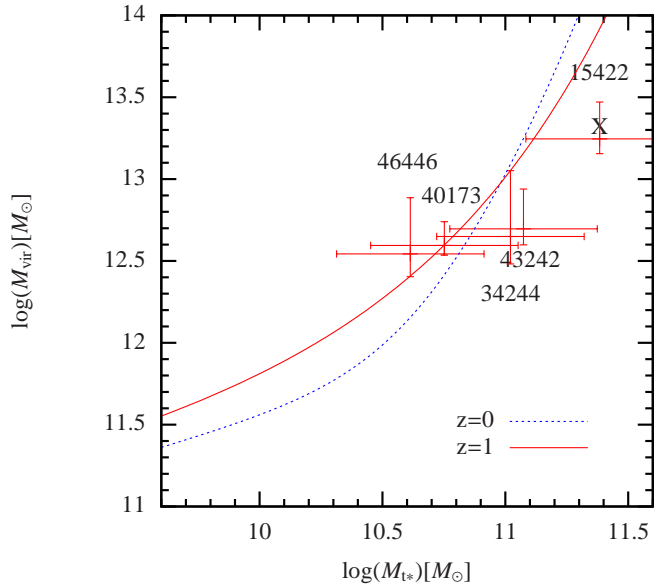}
\caption{\label{fig:hmsm} Halo mass estimated from lensing as a function of stellar mass. The dashed
and solid curves show the relation at redshift of 0 and 1, respectively, from
\citet{behroozi10}. The label ``X" shows the location of 15422 for the halo mass
calculated from the lens velocity dispersion of B04 model.} 
\end{center}
\end{figure}

The halo masses ($M_{\rm{vir}}$) of these galaxies thus obtained are shown as a
function of their stellar masses ($M_{{\rm t}\ast}$) in Fig.~\ref{fig:hmsm}. We
use total stellar masses\footnote{Note that the total stellar masses ($M_{{\rm
t}\ast}$) and stellar masses integrated inside the critical curve ($M_{{\rm
}\ast}$) are consistent within the 0.3~dex uncertainties.} from \cite{borc06}
which are given in Table~\ref{tab:sample}. The horizontal error bars show an
uncertainty of 0.3~dex in the stellar mass. The vertical bars
represent the masses within the 16$^{\rm{th}}$ and 84$^{\rm{th}}$ percentile of
the PDF. Note that the error bars on the mass do not include any systematic
uncertainties due to the assumption of an isothermal density profile
or the effects of adiabatic contraction of the halo. Observationally,
early-type galaxies show evidence for the contraction of the halo
\citep[see e.g.,][]{schulz10,dutton10} whereas late type
galaxies show the opposite \citep{dutton10}. Note that adiabatic
contraction, if present, will cause the density profile to be closer
to our assumption of an isothermal density profile in the inner
regions.

We also plot the halo mass$-$stellar mass relation inferred from the
subhalo-abundance matching technique by \citet{behroozi10} at redshift of 0
(dashed curve) and redshift of 1 (solid curve) in Fig.~\ref{fig:hmsm} for
comparison. Their relation at $z=0$ is in agreement with results from weak
lensing \citep{mandelbaum06}, kinematics of satellite galaxies \citep{more10},
and studies using a SDSS group catalog \citep{yang09} while the relation at
$z=1$ is consistent with results from \citet{drory09}. For 15422, we
also estimate the $M_{\rm{vir}}$ using the lens velocity dispersion of B04 and
total stellar mass from Table~\ref{tab:sample}. This is denoted by label ``X" in
Fig.~\ref{fig:hmsm} and is consistent within the uncertainties.

It is interesting that within the uncertainties of our measurements, all of the
five systems roughly follow the halo mass$-$stellar mass relation
predicted by the subhalo abundance matching technique. Although this does not
conclusively confirm the lensing nature of these candidates, it certainly adds
more credibility to the candidates being lens systems.

%%%%%%%%%%%%%%%%%%%%%%%%%%%%%%%%%%%%%%%%%%%%%%%%

\subsection{Cross-identification in Imaging at Other Wavelengths}
The {\HST} lens candidates from {\gems} were checked for
cross-identifications in data at other wavelengths, namely X-rays, mid-infrared, and
radio. The E-CDFS was observed with the Chandra X-ray telescope by
\citet{lehmer05}, with Multi-band Imaging Photometer for Spitzer (MIPS) at
24$\mu$m by \cite{papo04}, and the CDFS was surveyed in the radio using the Very
Large Array (VLA) by \citet{kellermann08}. None of our lens candidates were
found in X-ray imaging. However, astrometric comparison with the 24$\mu$m
and radio catalogs yielded one matched candidate.

Object 34244 from our sample is identified as object 82 from the VLA sample with
an offset of 0\farcs7 in position. The radio object 82 lies at an RA of $03^{\rm
h}32^{\rm m}06^{\rm s}.45$ and DEC of $-27^{\circ}47'29''.3$ and was measured to
have a flux density of 80~$\mu$Jy at 1.4~GHz.  This corresponds to a 1.4~GHz
rest-frame luminosity of 10$^{23.55}$~W~Hz$^{-1}$. This candidate is also
detected with MIPS at a flux density of 315.76 $\mu$Jy.

We inferred the total infrared (TIR) luminosity of 34244 both from the radio
flux using the radio-IR correlation \cite[e.g.,][]{yun01}, and from the
24~$\mu$m flux using the templates of \citet{char01} to convert it to TIR.
Hence, the TIR inferred from either the 1.4 GHz radio flux of 80$\mu$Jy
\citep{kellermann08} or the 24$\mu$m flux of 315.76 $\mu$Jy \citep{papo04}, is
$\sim$1--5 $\times 10^{11}$ $L_\odot$. This is in the range of the TIR
luminosity of the luminous infrared galaxies (LIRGs).  We use the radio--SFR and
TIR--SFR relations from \citet{bell03} to convert the radio and the TIR
luminosities into a star-formation rate.  The star-formation rates derived from
the radio and IR are 16 $M_\odot$/yr and 80 $M_\odot$/yr, respectively. Thus, in
accordance with their fluffy and not very sharp appearance the blue arc-like
features in 34244 could well be star-forming shells instead of being lensed
images of a background galaxy. We note that, recently, an optical spectrum of
the galaxy was taken with VIMOS on the Very Large Telescope (VLT) by
\citet{silverman10}. The slit had a width of 1~arcsec and was oriented along
north-south. The slit is expected to go through a part of the arc-like feature.
There is no clear detection of presence of a higher redshift object, however, any
signal in the composite spectrum could have been washed out due to the seeing
(J. Silverman, private communication). 

%
%%%%%%%%%%%%%%%%%%%%%%%%%%%%%%%%%%%%%%%%%%%%%%%%

\begin{table*}
\begin{center}
\caption{ Comparison of characteristics of recent lens surveys  \label{tab:surv}}
\begin{tabular}{cccccccc}
\tableline
\tableline
Name & Reference & Telescope & Bands  & Area & Depth &  lens & Confirmed\\ 
 &  &  &  &  (deg$^2$) & mag &  candidates & /followed-up systems\\ 
\tableline
\aegis              & \citet{moustakas06,moustakas07}      & \HST     &  $V$, $I$            & 0.18  & $V=28.75$ &   7  & 3/0    \\
\gems               & This work                            & \HST     &  $V$, $z$            & 0.22  & $V=28.25$ &  10  & 1/0    \\
\cosmos             & \citet{faure10}                      & \HST     &  $I$                 & 1.80  & $I=25.00$ &  88  & 4/18   \\
CFHTLS-{\sc sl2s}   & Gavazzi et al., in prep              & CFHT     &  $u$,$g$,$r$,$i$,$z$ & 170   & $g=25.47$ & 330  & 40/65  \\
SDSS-\slacs         & \citet{bolton08}                     & Sloan    &  spectroscopy        & 3732  & $r=17.77$ & 131  & 63/131 \\ 
\tableline
\end{tabular}
\end{center}
\end{table*}

\subsection{Future Investigations} 
In this paper, we have carried out various tests to investigate the lensing
nature of our candidates. However, none of our tests can be considered
conclusive and the candidates need further investigation. We give directions for
future work in order to confirm or rule out lensing and/or what one could learn
from these systems, if they are true lenses.

Although 15422 is a confirmed lens system with a known spectroscopic redshift for the
lens galaxy, no spectroscopic redshift for the background source is available.
We note that the lens subtracted images of B04 reveal another
source at a different redshift which is lensed as a quad with a smaller Einstein
radius than the arc visible in Fig.~\ref{fig:sample}. Spectroscopic redshifts of
both the background sources need to be measured for a detailed and sophisticated analysis of this
system. Currently, very few lens systems are known to have chance occurrence of
two background sources at different redshifts behind a galaxy-scale lens
\citep[e.g.,][]{gavazzi08,tu09}. These kind of systems are interesting because
one can constrain the slope of the mass density profile and cosmological
parameters to a better accuracy \citep{gavazzi08}.

The SFR calculations of the main galaxy in 34244 suggest that it is a
LIRG and the arc-like features could be sites of star formation. The orientation
of ellipticity from our mass models are misaligned to the light distribution.
Although such misalignments have been noted earlier, this might also imply that
34244 is less likely to be a lens system. Furthermore, an optical spectrum of
the candidate suggests no significant detection of a higher redshift object
which needs to be confirmed with a high resolution $-$ high SNR spectrum.

The current imaging and mass modelling are insufficient to comment on the
likelihood of lensing in 40173. It is not clear if the arc-like feature
corresponds to a background galaxy unless it is confirmed by spectroscopy.

The candidate 43242 shows two lines-of-sight to the same background galaxy
wherein one of the images can be seen rather unobstructed and the others are
seen through the dust in the foreground galaxy. Such lensed images can be used
to estimate the amount of differential dust extinction from the foreground lens
galaxy \citep{eliasdottir06}. This system could also be used to constrain the
disk mass in the lens galaxy which is a near edge-on spiral.

Although the model-predicted positions of lensed images in 46446 can be
reproduced and they show similar colors, the morphology is complex and perhaps
contaminated by dust in the lens galaxy. Spectroscopy and observations at other
wavelengths of this system can shed some light on the spectral type of the
candidate lens galaxy and confirm if it is a star-forming galaxy or not.

%%%%%%%%%%%%%%%%%%%%%%%%%%%%%%%%%%%%%%%%%%%%%%%%%%%%%%%%%%%%%%%%%%%%%%%%%%%%%%
\subsection{Comparison with other surveys} 
In this section, we briefly compare the efficiency of detection of lens systems
in \gems\ with that of the detection of lenses in other surveys.
Table~\ref{tab:surv} summarizes the characteristics of the surveys like \aegis,
\cosmos, CFHTLS-{\sc sl2s}, SDSS-\slacs\ along with the \gems\ survey. Lens
candidates in \aegis, \cosmos\ and \gems\ were selected from photometric data
from \HST\ while the candidates in CFHTLS-{\sc sl2s} were selected from
photometric data from ground-based telescope. SDSS-\slacs\, on the other hand,
used spectral information of galaxies from the SDSS and selected candidates
based on the presence of emission lines at redshifts higher than that of the
main galaxy. Another sample of ten lens candidates was reported from the \HST\
Medium Deep Survey \citep{ratna99}. However, we do not include their results in
our comparison because their sample is compiled from a set of 400 random fields
observed with varying limiting magnitudes which complicates the selection
function.

The number of lenses found in ground-based surveys is large thanks to their
large areal coverage. However, the small area, high resolution, deep surveys
carried out with the \HST\ are able to find a large number of lens candidates
per square degree. The exquisite resolution of these surveys allows them to find
lens systems with small image separation. These lens systems probe the lower end
of the halo mass function as opposed to the average lenses found in low
resolution wide-imaging surveys from the ground. Such systems are often missed
by ground-based surveys due to de-blending issues unless one uses spectral
information, as was done by the \slacs\ team. The depth of the \HST\ surveys
also allows detection of fainter lensed images around galaxies and also, lenses
that lie on average at higher redshifts.

Among the surveys carried out with the \HST, \cosmos\ has the largest area,
however it has imaging only in a single band. \gems\ and \aegis\ have a smaller
areal coverage but have imaging information in two distinct bands. The
multi-band information allows a more robust detection of lensed images with
distinct colors with respect to the lens galaxy. In this respect, the \gems\ and
the \aegis\ samples are expected to have a better completeness compared to \cosmos.
Also, the imaging for \cosmos\ was carried out in the $I$ band which is not
particularly suited for detecting lensed images given that they are usually
fainter in redder bands. On the other hand, in terms of the fidelity of the
samples, multi-band information is a double-edged sword. As can be seen from the
system 34244 in this paper, shells of star formation in the candidate lens
galaxy could also be confused to be lensed images thus affecting the purity of
the sample.

The \aegis\ survey has a very similar survey design in terms of area, depth and
wavelengths of imaging compared to \gems. Unless the surveys are severely
affected by cosmic variance, we expect to roughly find similar number of lenses
in these surveys. The visually identified candidate list from \aegis\ was
further matched to the spectroscopic information from the DEEP2 survey which
reduced false positives in their case.

We note that a quantitative comparison between the completeness, efficiency and
fidelity of lens samples in different surveys is difficult, primarily due to
inherent differences in the design of the surveys and methods used to select the
candidates. Large uncertainties in our current understanding of the faint end of
the galaxy luminosity function and the redshift distribution of sources
complicate the quantitative comparison further.

%%%%%%%%%%%%%%%%%%%%%%%%%%%%%%%%%%%%%%%%%%%%%%%%%%%%%%%%%%%%%%%%%%%%%%%%%%%%%%

\section{SUMMARY}
The \gems \ survey consists of \HST\ imaging in $V$ and $z$ bands within the
E-CDFS covering an area of 0.22~sq.~deg. We searched for lens candidates from a
sample of 25~000 galaxies in \combo\ \citep{borc06}. After applying our initial
selection criteria, a total of 1054 candidates were selected for visual
inspection and ranking. This resulted in ten lens candidates (that is, nine new
candidates and one known lens system) from the \gems\ survey, five of which are
best cases and were further investigated.

The lensing masses of these five candidates were calculated under the
assumptions of standard density profiles and a PDF for source redshifts. We
estimated stellar masses within the Einstein radii found from the lens models.
The stellar masses are consistently lower than or equal to the lensing masses,
as expected. We compared the projected dark matter fractions from galaxy-scale
lenses from surveys like \slacs\ and \cosmos\ to those in the \gems \ lens
candidates. Although uncertainty in the background source redshifts leads to
weak constraints on the estimated parameters, we find that all of the candidates
roughly follow the trend shown by lenses in \slacs\ and \cosmos. Also, these
candidates follow the halo mass$-$stellar mass relation of \citet{behroozi10}
within the uncertainties.

15422 is the only known and confirmed lens system in our sample. Our mass
estimate is consistent with that from B04. However, the imaging in B04 indicates
presence of a second lensed source in 15422 which deserves a detailed analysis. We
note that the main galaxy of 34244 is detected in radio, and the TIR luminosity
suggests that it is a LIRG and that the blue arc-like features might be
star-forming shells. The model ellipticity of the main galaxy of 46446 is
unusually high and inconsistent with the light distribution. Results from the
analyses of the candidates presented here suggest that 43242 is very likely
lens system, 46446 and 34244 are less likely to be lens systems whereas the
nature of 40173 is not clear. Spectroscopic observations of the candidates are
ultimately needed to definitively confirm or exclude lensing in these systems.
Also, redshifts of the background galaxies will allow the determination of the
mass of the lens more correctly. Our calculations imply that there should be
about eight lens systems in \gems. Given the large uncertainties in the
assumptions made in these calculations, it is broadly consistent with the two
good systems or ten plausible candidates in the sample.

%%%%%%%%%%%%%%%%%%%%%%%%%%%%%%%%%%%%%%%%%%%%%%%%%%%%%%%%%%%%%%%%%%%%%%%%%%%%%%
\acknowledgments
AM acknowledges support from CNRS and the ANR grant ANR-06-BLAN-0067. KJ is
funded by the Emmy Noether-Program of the German Science Foundation DFG. BH was
supported by STFC. The Dark Cosmology Centre is funded by the Danish National
Research Foundation. We thank Michael Cooper for providing the spectroscopic
redshifts for two systems beforehand. We are grateful to Mark Sargent for help
with the SFR calculations. We also thank David Hogg and Christine Ruhland for
interesting discussions. We appreciate the suggestions of the anonymous referee
which improved the content of the paper.

{\it Facilities:} \facility{HST (ACS)}, \facility{MPG/ESO:2.2m}
%
%%%%%%%%%%%%%%%%%%%%%%%%%%%%%%%%%%%%%%%%%%%%%%%%%%%%%%%%%%%%%%%%%%%%%%%%%%%%%%

\appendix
\section{FIVE LOW PROBABILITY CANDIDATES}
\begin{figure*} [h]
\begin{center}
\includegraphics[width=15cm]{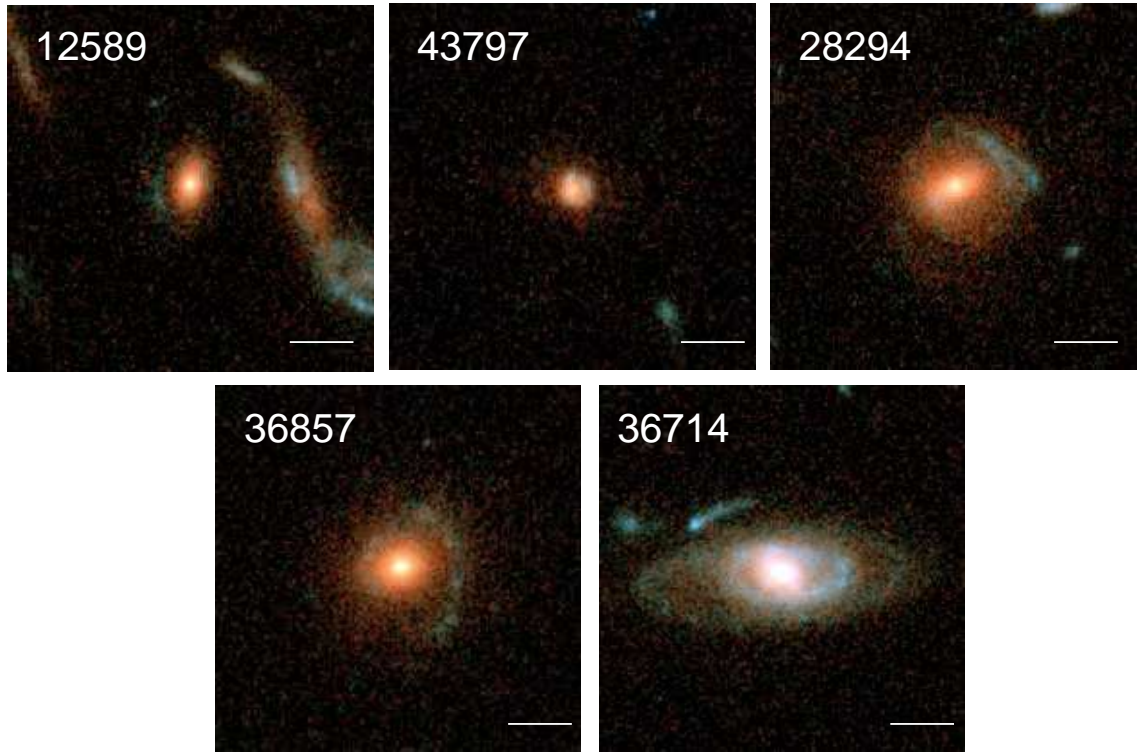} 
\end{center}
\caption{\label{fig:2ndsample} 
$V-z$ composite images of the lower ranked lens candidates of the \gems \ survey. 
Markers show a scale of 1\arcsec, the cutouts are roughly 6$\times$6 arcsec$^2$.
North is up and East is left.}
\end{figure*}

The five candidates with average ranks from 2 to 3 are presented here.  Color
images of these candidates are shown in Fig.~\ref{fig:2ndsample}. The candidates
are indicated with their identification number from \combo. Further information
like their positions, tile numbers, and photometric redshifts of the candidates
is given in Table~\ref{tab:2sample}. As before, the identification numbers and
the photometric redshifts are from \combo \ catalog. All candidates are on the
\gems \ tiles which range from 1 to 63. The $V$ and $z-$ band magnitudes in AB
are from \cite{cald08}. 

\begin{table*}
\begin{center}
\caption{Lower rank \gems \ candidates \label{tab:2sample}}
\begin{tabular}{ccccccc}
\tableline
\tableline
ID & RA & Dec & Tile & $V$ & $z$ & $z_{\rm {phot}}$\\ 
\tableline
12589 & 03:31:24.89 & $-$27:58:07.0 &  17  & 23.36 & 21.68 & 0.79  \\
43797 & 03:31:31.74 & $-$27:43:00.8 &  47  & 23.72 & 21.94 & 1.02  \\
28294 & 03:31:50.54 & $-$27:50:28.4 &  33  & 22.50 & 20.32 & 0.66  \\
36857 & 03:31:53.24 & $-$27:46:18.9 &  38  & 22.19 & 20.44 & 0.42  \\
36714 & 03:32:59.78 & $-$27:46:26.4 &  37  & 20.92 & 19.73 & 0.44  \\
\tableline
\end{tabular}
\end{center}
\end{table*}

%%%%%%%%%%%%%%%%%%%%%%%%%%%%%%%%%%%%%%%%%%%%%%%%%%%%%%%%%%%%%%%%%%%%%%%%%%%%%%

\end{document}